\documentclass[aps,prl,twocolumn,superscriptaddress,showpacs,floatfix]{revtex4-1}

\usepackage{graphicx}
\usepackage{color}
\usepackage{tabularx} 
\usepackage{bm} 

\usepackage[normalem]{ulem}
\usepackage{amsmath,amssymb}

\begin{document}

\title{Magnetic Trapping of Cold Methyl Radicals}
\author{Yang Liu}
\author{Manish Vashishta}
\author{Pavle Djuricanin}
\author{Sida Zhou}
\affiliation{Department of Chemistry, The University of British Columbia, Vancouver, BC V6T 1Z1, Canada}
\author{Wei Zhong}
\affiliation{Department of Physics and Astronomy, The University of British Columbia, Vancouver, BC V6T 1Z1, Canada}
\author{Tony Mittertreiner}
\affiliation{Department of Chemistry, The University of British Columbia, Vancouver, BC V6T 1Z1, Canada}
\author{David Carty}
\affiliation{Durham University, Joint Quantum Centre Durham - Newcastle, Departments of Chemistry and Physics, South Road, Durham DH1 3LE, UK}
\author{Takamasa Momose}
\affiliation{Department of Chemistry, The University of British Columbia, Vancouver, BC V6T 1Z1, Canada}
\affiliation{Department of Physics and Astronomy, The University of British Columbia, Vancouver, BC V6T 1Z1, Canada}
\email{momose@chem.ubc.ca}
\date{\today} 
 \begin{abstract}
We have demonstrated that a supersonic beam of methyl radicals (CH$_3$) in the ground rotational state of both $para$ and $ortho$ species has been slowed down to  standstill  with a magnetic molecular decelerator, and successfully captured spatially in an anti-Helmholtz  magnetic trap  for $>$ 1 s.  The  trapped CH$_3$ radicals have a mean translational temperature of  about 200 mK with an estimated density of $>5.0\times10^7$  cm$^{-3}$.  The methyl radical is an ideal system for the study of cold molecules not only because of its high reactivities at low temperatures, but also because further cooling below 1 mK is plausible via sympathetic cooling with ultracold atoms. The demonstrated trapping  capability of methyl radicals opens up various possibilities for realizing ultracold ensembles of molecules towards Bose-Einstein condensation of polyatomic molecules and investigations of reactions governed by quantum statistics.\end{abstract}

\date{\today}
\maketitle

Since the realization of Bose-Einstein condensation (BEC) of ultra-cold atoms \cite{Anderson:1995dw,Davis:1995bj}, experimental observation of correlated motion of matter waves of particles in an ensemble has drawn considerable attention in various fields, revealing new quantum aspects of matter. 
Especially, the research of cold and ultracold molecules has expanded rapidly over the past decade because of their importance in various fields from fundamental physics to interstellar chemistry \cite{ColdMolecules,Carr2009,Jin2012}.   

Recent developments in molecular deceleration \cite{vandeMeerakker:2012ft} now make it possible to control the translational motion of molecules and create extremely slow molecular beams. 
Among various molecules, free radicals having unpaired electron(s) are of interest in relation to cold scattering because of their high reactivity towards other molecules and substances. 
Since molecules with unpaired electron(s) always have a non-zero magnetic moment, it is natural to use magnetic fields for the deceleration, i.e.~Zeeman deceleration \cite{Vanhaeck2007,Hogan_2008,Narevicius2008Ne,Narevicius:2008dv,Wiederkehr2012,Liu2014},  and as well as for trapping \cite{deCarvalho1999}  of these reactive species. 
So far, only two diatomic molecules, He$_2$ \cite{Motsch_2014,Jansen_2015} and O$_2$ \cite{Narevicius:2008dv,Wiederkehr2012,Liu2014}, have been successfully decelerated by Zeeman deceleration, except for our preliminary work on methyl radicals (CH$_3$) \cite{Momose:2013hy}. 
The magnetic trapping of CaH \cite{Doyle:1998di}, CaF \cite{Lu:2014dq}, NH \cite{Tsikata2010,Riedel:2011kk}, and OH \cite{Sawyer:2007ga} were reported following buffer gas cooling, and optical loading as well as Stark deceleration.  The trapped molecules were applied to high-resolution spectroscopy \cite{Friedrich:1999bf},  cold collision experiments \cite{Sawyer:2008fl} as well as evaporative cooling \cite{Stuhl:2013bh}. However, all the molecular candidates that have been magnetically decelerated and trapped have been limited to diatomic species.  Here, we report the magnetic trapping of CH$_3$, the simplest organic polyatomic radical, in the ground rotational state for the first time for longer than a second, after slowing down its velocity by well-controlled pulsed magnetic fields.

Creating cold ensembles of the methyl radical is of particular importance.   CH$_3$ is an ideal system for long trapping experiments, because CH$_3$ does not posses an electric dipole moment \cite{Yamada1981},  and therefore there is no trap loss for CH$_3$ due to rotational excitation by blackbody radiation \cite{Hoekstra:2007dk}. Furthermore, it has been predicted that sympathetic cooling of CH$_3$  with ultracold ($<$ 100 $\mu$K) alkaline-earth metals may be possible, thereby enabling the further cooling of CH$_3$ into the ultracold regime  ($<$ 1 mK) \cite{Tscherbul:2011eh}.  Achieving long trapping times and lower temperatures, by sympathetic cooling, are both imperative for the realization of molecular BEC.
  Another key feature of CH$_3$ is that it exists as two spin isomers, {\it para} ($I$ = 1/2) and {\it ortho} ($I$ = 3/2), which is an ideal system to solve the outstanding problem of understanding the rate and mechanism of nuclear spin conversion between spin isomers \cite{Miyamoto:2008cha} as well as the separation of different nuclear spin states of molecules \cite{Hougen:2005dd,Sun:2005em,Kravchuk319}.

\begin{figure*}[ht!]
\begin{center}
 \includegraphics[bb = 0 0 1379 556, width=0.8\textwidth]{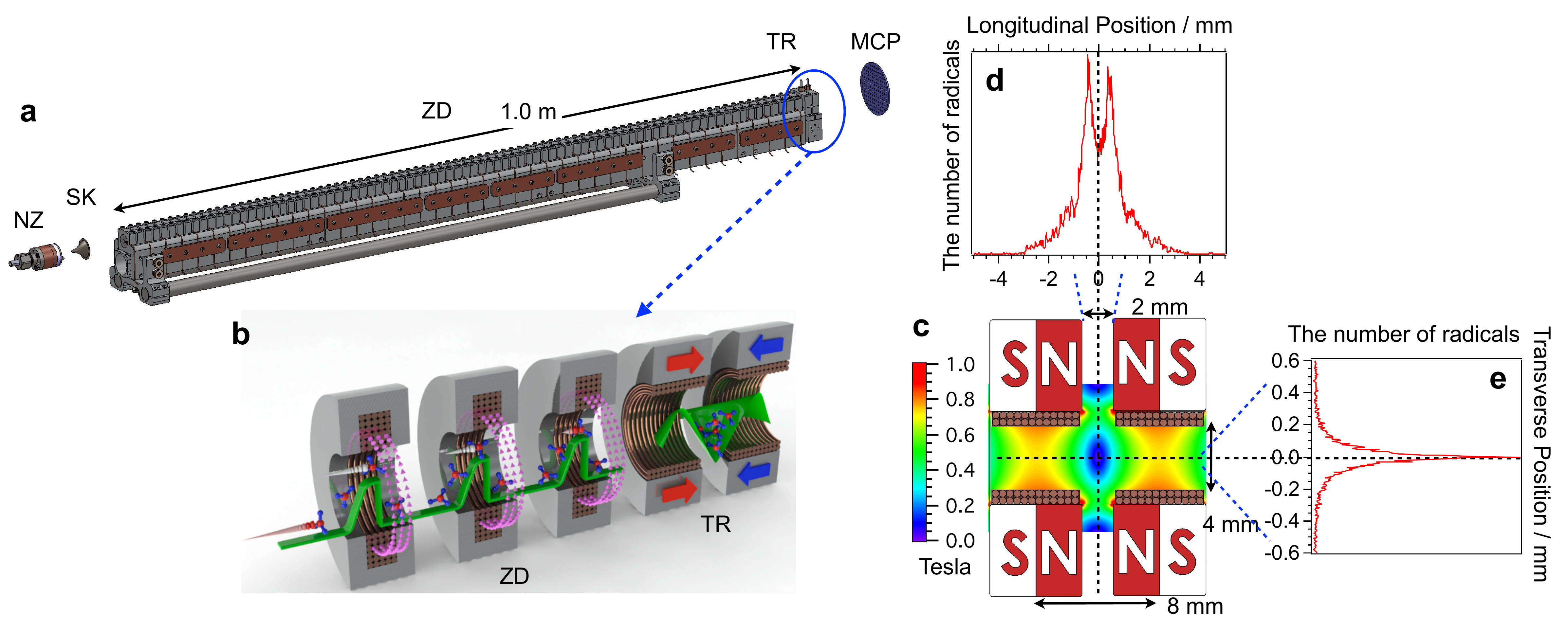}
\caption{ (a) The system consists of a cold pulsed discharge nozzle (NZ), a skimmer (SK), a 85-stage Zeeman decelerator (ZD), an anti-Helmholtz magnetic trap (TR), and a micro-channel plate (MCP) detector. 
(b) Cross section of the last three decelerator coils (ZD) and the trap magnets (TR). The purple triangles of each decelerator coil show the direction of  current flow. The bold arrows of the trap magnets show the direction of the magnetic fields. The green lines inside the coils and the trap magnets show an effective potential for the CH$_3$ radicals traveling through the center of the decelerator.
(c) The cross section of the magnetic field distribution inside the trap. The dimensions are also shown in the figure. A solenoid coil with an inner diameter of 4 mm was placed inside each magnet. The spacing of two magnets was 2 mm. (d) A snapshot of the spatial distribution of the radicals along the longitudinal axis obtained by a simulation. A distribution at 100 ms after the trapping is shown as an example.  Distributions at different times are very similar to this.
(e) A snapshot of the transverse distribution of the radicals.}
\label{Fig1}
\end{center}
\end{figure*}

Figure 1a shows  a schematic of the  decelerator and trap used in the present experiment.
The decelerator was a 1m-long linear magnetic decelerator comprised of 85 solenoid coils \cite{Vanhaeck2007,Narevicius2008Ne,Liu2014}. 
Each coil was comprised of 30 turns of Cu wire with an inner diameter of 4 mm and a length of 3 mm. 
A current of 600 A creates a maximum field of 4.27 T at the center of the coil, which provides a potential energy of  $\pm$ 2.0 cm$^{-1}$ to CH$_3$ \cite{Momose:2013hy}.  
The complete coil assembly as well as all copper wires were maintained at 110 K with a closed-cycle refrigerator in order to remove the heat caused by the high current pulses, and also  to decrease the resistance of the Cu wires.
The translational velocity of radicals in low-field seeking (LFS) states can be permanently reduced by switching off the magnetic field within 3.5 $\mu$s before the molecules reach the downhill section of the potential (Fig.~1b). Repeating the deceleration and switching process for a series of solenoid coils can gradually bring the molecules to a standstill.

In order to trap the decelerated radicals, an anti-Helmholtz type magnetic trap made of two ring-shaped NdFeB permanent magnets was placed at the end of the decelerator  (Fig.~1b). Its magnetic field distribution is shown in Fig.~1c. This trap provides a maximum magnetic field of 0.75 T (longitudinal) and 0.45 T (transverse), which corresponds to a 500 mK (longitudinal) and 300 mK (transverse) trap depth for CH$_3$. In addition, solenoid coils were placed inside the magnets and worked as the last stages of the decelerator \cite{Hogan_2008,Wiederkehr2010,Akerman2016}. 
  
 Figures 1d and 1e show a snapshot of the spatial distribution of the trapped radicals inside the magnetic trap along the longitudinal and transverse directions, respectively, as predicted by particle simulations. Although the maxima of the magnetic fields along the longitudinal direction are separated  by 8 mm, 75 \% of the radicals are confined within $\pm$ 1 mm at the centre of the trap, and 95 \% are within $\pm$ 2 mm. On the other hand, about 90 \% of the radicals are confined within $\pm$ 300 $\mu$m along the transverse direction due to the focusing effect of our Zeeman decelerator coils \cite{Liu2014}. 
  
CH$_3$ molecules were generated by a discharge of a gas mixture of 15\% CH$_4$ seeded in Kr. By cooling a home-made pulsed nozzle down to 150 K with liquid N$_2$, a mean velocity of 340 m s$^{-1}$  was obtained for a free flight beam. 
The detection of the radical was done with  2+1 resonance enhanced multi photon ionization (REMPI) spectroscopy via the $4p$  Rydberg state \cite{Black1998} using nano-second pulsed radiation at 286.3 nm (Sirah Precision Scan, 3.2 mJ, 13 ns pulse width).  The laser pulses were focused at the center of the trap with a $f=300$ mm lens through the 2 mm slit between the two trap magnets.  The beam waist at the focal point was 40 $\pm$1.3 $\mu$m as determined by a standard knife edge measurement. The ionized radicals were accelerated by an electric field applied between two plates inside the trap (not shown in Fig.~1b)  along the radical beam axis and detected by a micro-channel plate (MCP) detector placed behind the trap.

By adjusting the magnetic field strength produced by the solenoid coils along with the motion of the molecules, a portion of the radicals in the LFS states can be slowed to as low as 20 m s$^{-1}$. 
REMPI frequency spectra \cite{Black1998} showed transitions from $N''$ = 0 ($S$(0)) and $N''$ = 1 ($R$(1) and $S$(1)) states of CH$_3$, but no transitions from rotational states higher than $N''$ = 1 such as $P$(2), $O$(2), or $R$(3) as shown in Fig.~2. These spectra indicate that the rotational cooling of CH$_3$ radicals during the supersonic expansion was efficient enough to produce only the lowest rotational states of each nuclear spin isomer of {\it ortho} ($N''$ = 0, $K''$ = 0) and {\it para} ($N''$ = 1, $|K''|$ = 1) selectively. It should be noted that both states are simultaneously decelerated in this experiment (Fig.~2 bottom trace), because the Zeeman shift of these two states are almost identical \cite{Momose:2013hy}.

\begin{figure}[t!]
\begin{center}
  \includegraphics[bb = 0 0 494 398, width=0.38\textwidth]{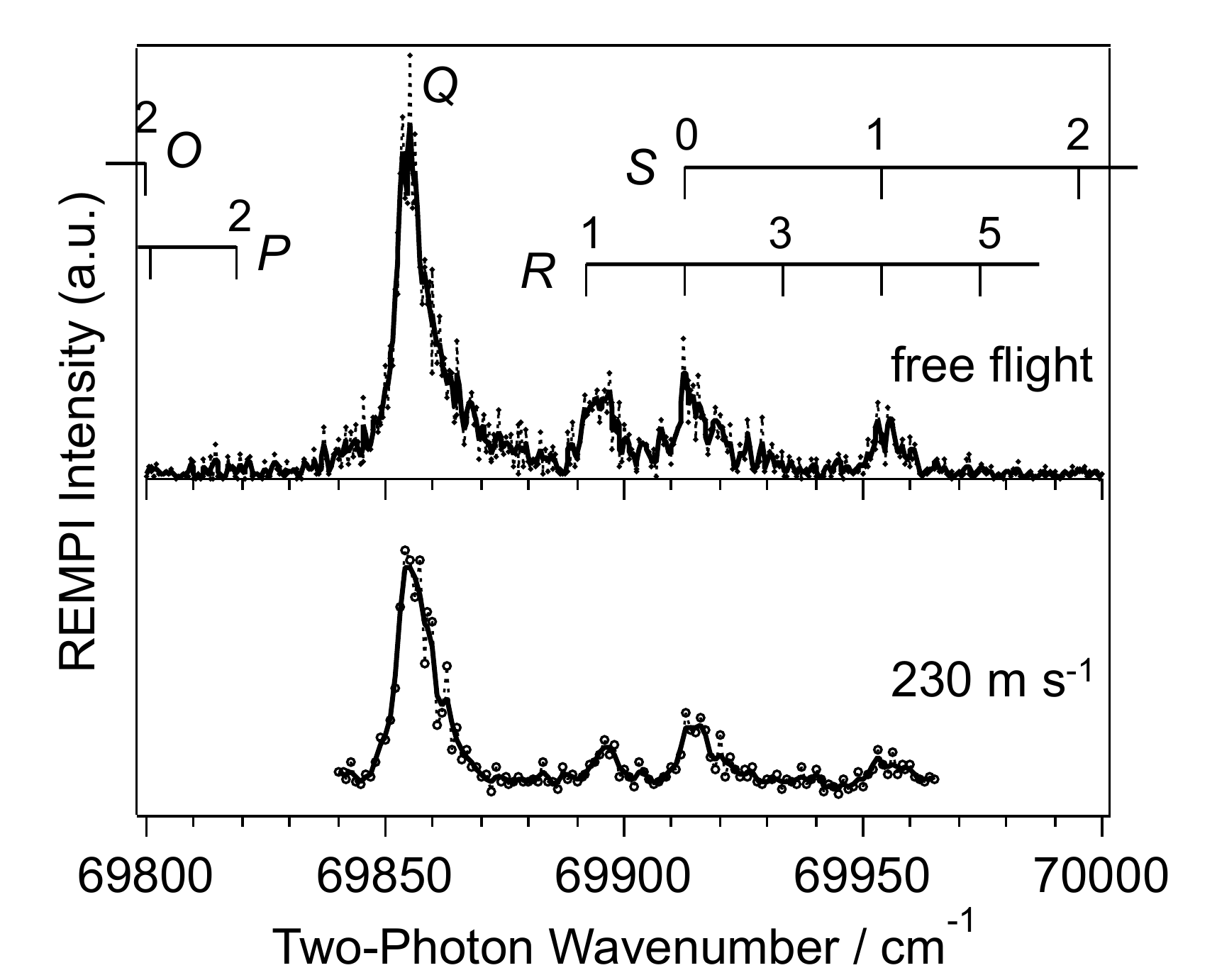}
\caption{A REMPI frequency spectrum of CH$_3$ in a free flight beam (top) and in a decelerated beam (bottom, the mean velocity was 230 m s$^{-1}$). Spectroscopic assignments are given on top of each peak in the top trace with $O$, $P$, $Q$, $R$, and $S$ corresponding to $\Delta N$ = $-2$, $-1$, $0$, $1$ and $2$, respectively.}
\label{Fig1}
\end{center}
\end{figure}

For trapping experiments, the decelerator was configured such that a target CH$_3$ radical had a longitudinal velocity of about 60 m s$^{-1}$ before entering the trap region. The last two coils placed inside the magnets were utilized to bring the target CH$_3$ radical packet to a complete standstill before it reached the middle of the rear magnet. Upon termination of the current flow in the two solenoids, complete trapping of the radical packet between the two permanent magnets was achieved.

\begin{figure}[t!]
\begin{center}
 \includegraphics[bb = 0 0 491 582, width=0.38\textwidth]{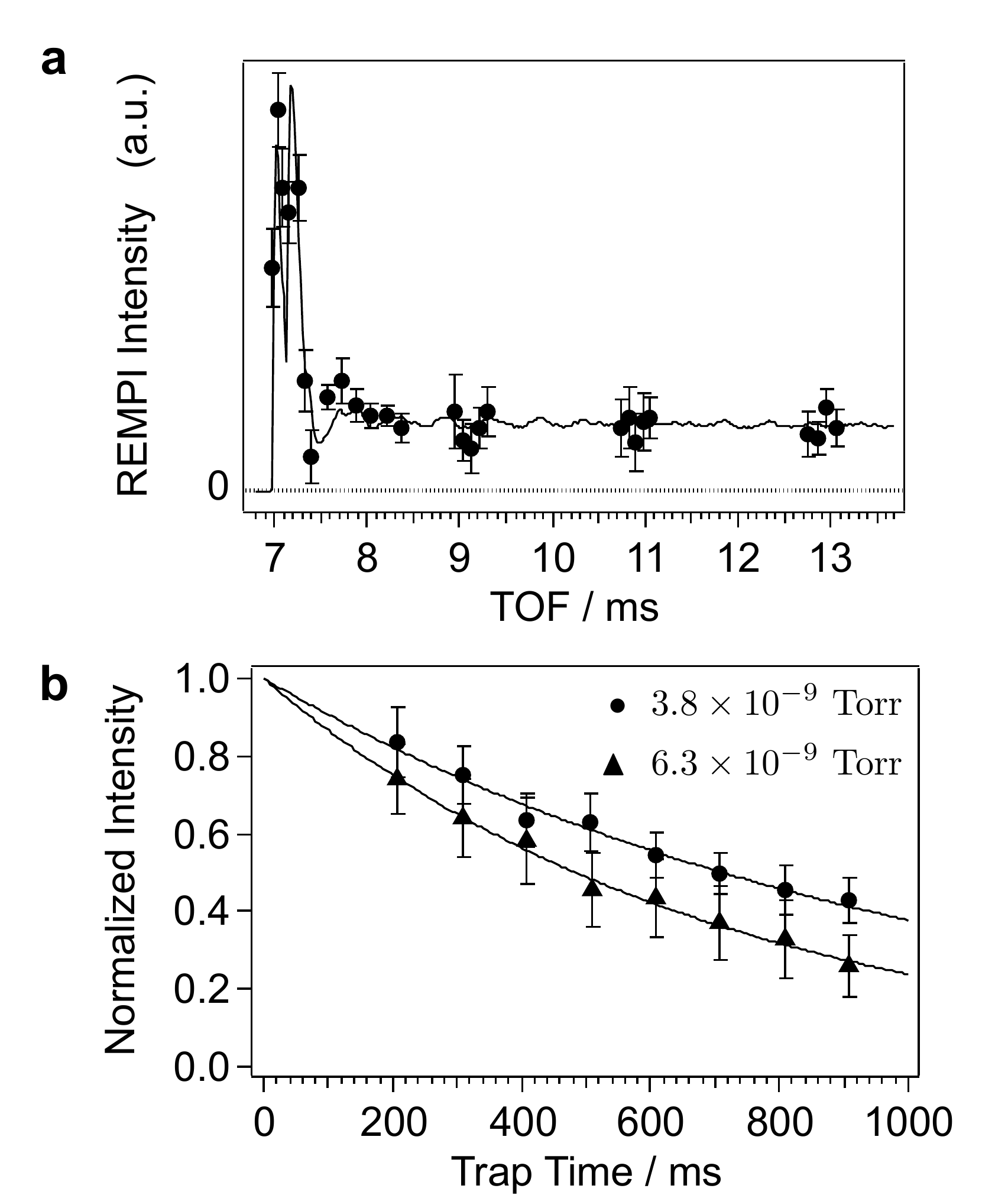}
\caption{(a) Observed TOF (filled circle) of trapped CH$_3$.  The experimental data shown is an average of 128 sets of subtracted signal intensities between the Zeeman decelerator on and off, which were executed alternatively in each sequence. The zero-intensity line (baseline) is shown as a dotted line. 
The solid trace shows a simulated TOF feature.  (b) Trapped signal intensity over 1 s normalized to the intensity at TOF= 10 ms for each fitted curve.  The solid lines show fitted curves with a single exponential decay function for each pressure.}
\label{Fig3}
\end{center}
\end{figure}

Figure 3a shows a time-of-flight (TOF) feature of the REMPI signal of CH$_3$ at the centre of the trap. In order to eliminate background signals and other noise sources effectively, experiments with the Zeeman decelerator on and off (i.e.~no deceleration) were executed alternatively in each sequence, and the on and off signal intensities of each set were subtracted before taking their averages.  The zero-intensity line shown in Fig.~3a (the dotted line) corresponds to the case with no radicals in the trap and is the measured baseline by definition, while positive values indicate that some radicals decelerated by the decelerator were detected inside the trap.   As shown in Fig.~3a, a strong REMPI signal was observed at TOF =7.0 ms $-$ 7.5 ms, which corresponds to the first entry of a dense decelerated radical packet into the trap region. After that, signals were detected continuously, which confirms that  the decelerated CH$_3$ radicals were successfully confined in the trap. The observed TOF feature was well reproduced by trajectory simulations of the radicals inside the trap, which is shown as a solid trace in Fig.~3a.
 
The trapped signal showed an exponential decay over time for 1 s. The two traces shown in Fig.~3b, which were recorded every 100 ms with a 10 Hz pulsed laser, correspond to the cases where the background pressure was $3.8\times10^{-9}$ Torr (filled circle) and $6.3\times10^{-9}$ Torr (filled triangle). The fitted $1/e$ lifetime, $\tau$, of the trapped radicals was $\tau = 1.03\pm 0.19$ s and $0.70\pm 0.16$ s, respectively.

The shorter trap time at higher background pressure indicates that the trap lifetime of CH$_3$ is limited by collisions with residual background gases. Under such conditions, the number of trapped radicals, $N_{{\rm CH_3}}$, may obey a first order differential equation $dN_{{\rm CH_3}}/dt = -(1/\tau)N_{{\rm CH_3}}$, where the lifetime, $\tau$, is inversely proportional to the average of the product of the cross section, $\sigma$, velocity, $v$, and the density, $n$, of background gas particles \cite{VanDongen:2011gg}. The measurement of background gases by a residual gas analyzer revealed that the major component of the background gas in our vacuum chamber was H$_2$ (or H) with some N$_2$ as a minor component (less than 1/3 of H$_2$). By assuming that the trap loss is entirely induced by H$_2$ molecules ($v$=1950 m s$^{-1}$ mean velocity at 300 K), the average of the cross section for the trap loss caused by H$_2$ molecules is obtained to be $\langle\sigma\rangle _{{\rm H_2}}$ $\sim$ 380 \AA $^2$ at the present trap depth of 300 $\sim$ 500 mK. 

The cross section obtained here is an effective cross section that depends on the trap depth and the velocity of the collision partners. This value is necessarily less than but within an order of magnitude of the total cross section due to the finite trap depth.  From previous works on trap loss rates of cold atoms due to collisions with foreign gases \cite{VanDongen:2011gg,Lam:2014fc}, it is known that the trap loss rate varies quite strongly with trap depth especially near the energy scale for quantum diffractive collisions, $\epsilon_d=4\pi \hbar^2/m\sigma_T$,  where $m$ is the mass of trapped particles and $\sigma_T$ is the total collision cross section  \cite{Bali99}. This energy scale is on the order of $\sim$ 100 mK for the present case.  Therefore, our measured rate is at least a few times  smaller than the total collision rate. Further cooling is indispensable in order to obtain the cross section at the limit of zero potential depth. 
It is also noted that the cross section obtained here is an average of those of {\it ortho} and  {\it para}  CH$_3$. The cross section of these two species must be different due to the difference in the interaction; the interaction in the   {\it para} species is more anisotropic than that in the  {\it ortho} species.  Further improvement in our sensitivity to detect each spin component separately is underway in order to reveal the role of anisotropy of the interaction potential in molecular collisions.

The number of trapped radicals and its density can be estimated from the efficiency of the 2+1 REMPI process with the help of simulations. The  2+1 REMPI process of CH$_3$ is less efficient than other molecules due to  predissociation at the intermediate $4p$ Rydberg state \cite{Black1998}.   Indeed, the REMPI spectra shown in Fig.~2 has a linewidth of 4.5 cm$^{-1}$ in the Q branch, which must be limited by the predissociation lifetime. A  predissociation rate of $8.3\times 10^{11}$ s$^{-1}$ estimated from the linewidth is faster than the ionization rate of  $7.0\times10^{10}$ s$^{-1}$, assuming a standard ionization cross section of $\sigma^{(i)}=1.0\times10^{-17}$ cm$^2$ and a photon flux of $F=7.0\times10^{27}$ cm$^{-2}$s$^{-1}$ of our pulse.  By using a steady state approximation \cite{Bo1989,Zhou:2003en} with a two-photon excitation cross section of  50 GM (=$5 \times 10^{-49} $ cm$^4$s) and a 10 $\%$ quantum efficiency for the MCP detector to detect positive ions, signal intensities of the ions are predicted to be  0.0022 $N$, where $N$ is the number of radicals in the ground state.  This estimation indicates that  450 radicals are necessary in the detection volume (a cylinder of 40 $\mu$m radius and 600 $\mu$m length) to obtain one signal count at the MCP. 
Furthermore, our simulation predicts that about 0.8 \% of the  trapped radicals reside inside the detection volume.  Therefore, the total number of trapped radicals is estimated to be 55000, which is equivalent to a density of $5.0\times10^7$  cm$^{-3}$ in the effective trap volume (a cylinder of 300 $\mu$m radius and 4 mm length).  

The above estimation is a lower bound for the trapped radicals. The two-photon cross section of 50 GM is a good approximation for molecules with a large two-photon cross section \cite{ChandraJha:2008dl} and therefore it might be overestimated for CH$_3$. 
It is noted that the peak density of the trapped OH radicals in Ref.~\cite{Stuhl:2013bh} was determined to be $5\times10^{10}$ cm$^{-3}$, in which  the production rate of the radicals as well as the efficiency of the deceleration are roughly similar to our experiments.

Our simulation indicates that the velocity deviation of CH$_3$ in the present trap is about $\pm 10$ m s$^{-1}$, corresponding to a translational temperature of 200 mK (FWHM) in the trap.  
One immediate application of the trapped cold radicals is precision spectroscopy \cite{Jansen_2015} for the determination of molecular parameters of the radicals precisely by resolving  hyperfine structures  in  infrared transitions \cite{Davis:1997bx}.
Another application is the observation of cold reactive collisions. The thermal de Broglie wavelength of CH$_3$ at this temperature is 1.0 nm, which is already a few times larger than the classical molecular size of CH$_3$ (0.4 nm). 
A deviation in the collisional behavior from high-temperature collisions might  be detected with this fundamental intermediate in hydrocarbon chemistry, which will shed light  on reactions at low temperature  environments such as planetary atmospheres and the interstellar medium \cite{CA1975,Jasper:2007fb}.   



The study was supported by a National Science and Engineering Research Discovery Grant in Canada and funds from Canada Foundation for Innovation for the Centre for Research on Ultra-Cold Systems (CRUCS) at UBC. The visit of David Carty was supported by the EPSRC Programme Grant MMQA: MicroKelvin Molecules in a Quantum Array (EP/I012044/1).


%

\end{document}